\documentclass[a4paper]{article}

\usepackage[textsize=tiny]{todonotes}
\usepackage{microtype}
\usepackage{graphicx}
\usepackage{subfigure}
\usepackage{breqn,bm}
\usepackage{lipsum}

\usepackage{booktabs} 

\usepackage{xcolor} 
\usepackage{algorithm}
\usepackage{algorithmic}
\usepackage{eso-pic} 
\usepackage{forloop}
\usepackage{url}

\usepackage{amsmath}
\usepackage{amssymb}
\usepackage{mathtools}
\usepackage{amsthm,nicefrac}

\usepackage[capitalize,noabbrev]{cleveref}

\theoremstyle{plain}
\newtheorem{theorem}{Theorem}[section]

\newtheorem{lemma}[theorem]{Lemma}

\theoremstyle{definition}
\newtheorem{definition}[theorem]{Definition}
\newtheorem{assumption}[theorem]{Assumption}

\theoremstyle{remark}

\allowdisplaybreaks

\usepackage{INTERSPEECH2022}[accepted]

\title{SepIt: Approaching a Single Channel Speech Separation Bound}
\name{Shahar Lutati$^1$, Eliya Nachmani$^{1,2}$, Lior Wolf$^1$}
\address{
  $^1$Tel-Aviv University\quad $^2$Meta AI Research}
\email{shahar761@gmail.com, enk100@gmail.com, wolf@cs.tau.ac.il}

\begin{document}

\maketitle
\begin{abstract}
We present an upper bound for the Single Channel Speech Separation task, which is based on an assumption regarding the nature of short segments of speech. Using the bound, we are able to show that while the recent methods have made great progress for a few speakers, there is room for improvement for five and ten speakers. We then introduce a Deep neural network, SepIt, that iteratively improves the different speakers' estimation. At test time, SpeIt has a varying number of iterations per test sample, based on a mutual information criterion that arises from our analysis. In an extensive set of experiments, SepIt outperforms the state of the art neural networks for 2, 3, 5, and 10 speakers.
\end{abstract}
\noindent\textbf{Index Terms}: speech separation, single channel, deep learning

\section{Introduction}
Single Channel Speech Separation (SCSS) problem is a specific setting of the general Blind Source Separation (BSS) problem. By employing deep neural networks, great improvement has been achieved in the last few years in separating two and three speakers~\cite{luo2020dualpath,nachmani2020voice}, leading to the very recent state of the art \cite{subakan2021attention}. Somewhat disappointingly, in terms of model sizes versus improvement obtained, the model size has more than tripled ($26M$ \cite{subakan2021attention} vs $7.5M$ \cite{nachmani2020voice}), while the improvement as measured when training on same dataset without dynamic mixing is only $0.1dB$. One may, therefore, wonder whether additional improvement in performance is possible. 
This drives the need for a theoretical upper bound for SCSS. 

SCSS has unique characteristics in comparison to other BSS problems. The speech signal is not stationary, unless short segments are considered. Jensen et al. \cite{Jensen} have shown empirically that when the length of the speech segment is longer than $20[ms]$, the distribution is closer to the Laplace distribution than to the normal distribution. As a result, the known bounds for BSS~\cite{4202618} do not hold. In this work, we derive such a bound by employing the assumption of a Laplacian distribution. Treating the speech mixture as a random process, we derive a bound that expresses the maximum achievable Signal to Distortion Ratio (SDR) by any neural network. We then present a deep learning method called SepIt, which uses the bound during its training. 

In an extensive set of experiments, we validate the assumption made in our analysis, as well as the bound itself. We then show that the SepIt model outperforms the state of the art in separating two and three speakers for the WSJ \cite{garofolo1993csr} benchmark and five and ten speakers for the LibriMix \cite{cosentino2020librimix} benchmark.

\smallskip
\noindent{\bf Related Work\quad} SCSS using deep learning techniques has been explored intensively in the last years. Erdogan et al. \cite{7178061} introduced a phase-sensitive loss function trained using a LSTM neural network. Hershey et al. \cite{hershey2016deep} developed the WSJ-2mix benchmark and proposed a neural separation network with a clustering-based embedding. The TasNet architecture \cite{luo2018tasnet} employs a time domain encoder-decoder. Subsequently, Luo et al. \cite{luo2019conv}  introduced the ConvTasNet architecture based on a convolutional neural network with masking. ConvTasNet was further improved with the dual-path recurrent neural network (DPRNN) architecture by \cite{luo2019dual}. Zeghidour et al. \cite{zeghidour2020wavesplit} introduced Wavesplit, which uses clustering on speaker representation to separate the mixture. Nachmani et al. \cite{nachmani2020voice} proposed the VSUNS model, which is based on the DPRNN model, but removes the masking sub-network. The SepFormer architecture is a transformer-based architecture that captures both short- and long-term dependencies \cite{subakan2021attention}. Yao et al. \cite{yao2022stepwise} introduced a coarse-to-fine framework called the Stepwise-Refining Speech Separation Network (SRSSN). Hu et al. \cite{hu2021speech} presented the Fully Recurrent Convolutional Neural Network (FRCNN) architecture, which uses lateral connections to fuse information processed at various timescales. 

Independently, upper bounds have been developed for the BSS problem. Sahlin et al. \cite{534883} derived an asymptotic Cramer Rao bound for the MIMO case, which assumes a full-rank channel matrix. Doron et al. \cite{4202618} presented a Cramer Rao bound for Gaussian sources. Kautsky et al. \cite{2020} perform an analysis for both the Gaussian and non-Gaussian sources, presenting a Cramer Rao induced bound. However, the analysis only addresses linear operations over the mixture signal. As far as we can ascertain, no upper bound for non-Gaussian sources exists for non-linear methods such as Deep Neural Networks.

\section{SCSS Upper bound}
\label{sec:upper_bound}

We start by formulating the required lemmas and stating the assumption we employ. 
\vspace{-2mm}
\begin{lemma}[Fisher Information Upper Bound]
From \cite{Brunel} the Fisher Information $J(\theta)$ is upper bounded by the Mutual Information $I(X,\theta)$:
\begin{equation}
    J(\theta) \leq I(X,\theta)
\end{equation}
\label{lemma:fisher_bound}
\end{lemma}
\vspace{-6mm}
Where $\theta, X$ are random variables as defined in \cite{Brunel}.

\begin{lemma}[Joint Estimation Lemma]
For $N$ i.i.d random variables with parameter $\theta$, the joint Fisher Information is
\begin{equation}
    J(\theta_{joint}) = N\cdot J(\theta)
\end{equation}
\label{lemma:joint_fisher}
\end{lemma}
\vspace{-3mm}
Jensen et al. \cite{Jensen} demonstrated empirically that speech, divided into short segments, neglecting quiet segments, is well modeled as a stationary process. Furthermore, short segments of a single speaker follow a Laplace random variable. 
\vspace{-2mm}
\begin{assumption}
Short speech segments can be captured by a Laplace random variable with zero mean and scale parameter $b$.
\begin{equation}
    f(voice = x) = {\nicefrac{1}{2}b^{-1}}{e^{-|\frac{x}{b}|}}
\end{equation}
\label{asmp:laplace}
\end{assumption}
\vspace{-4mm}


\noindent \textbf{Mixture Model\quad }
Let the different speakers' signals in the mixture compose a matrix $V=[v_0,v_1,...,v_{C-1}]^T,$
where $v_i \in \mathbb{R}^{L}$ is a single speaker signal in the time domain, with length $L$. To maintain the stationary requirements we use windowing of size $w$ as in Assumption ~\ref{asmp:laplace}.
The mixture $m = \sum_i a_i\cdot v_i$ is modeled as a linear combination of the columns of $V$, 
where $a_i\in\mathbb{R}$ are the mixture coefficients.
\begin{definition}[Signal to Distortion Ratio (SDR)]
Let the error $\epsilon$ be defined as $\epsilon =  v_0 - \bar{v}_0$.
The Signal to Distortion Ratio (SDR) is,
\begin{equation}
    SDR = 10log_{10}\frac{Var(v_0)}{Var(\epsilon)}
\end{equation}

\label{defintion:sdr}
\end{definition}
Denote a single segment of size $w$ from $m$ as $m_r$, and a single segment of $v_i$ as  $v_{i,r}$, where $0\leq r \leq \frac{L}{w}$. Without loss of generality, assume that the goal is to separate $v_0$ from the mixture $m$. Denote by $\bar{v}_0$ the estimation of $v_0$, using deep neural network, $D$.





\begin{lemma}
Given a single segment $v_{0,r}$,$m_r$ and deep neural network, $D$, the following inequality holds,
\begin{equation}
    \frac{L}{w} \cdot I(m_r,v_{0,r}) \geq J(\bar{v}_{0})
\end{equation}
\end{lemma}
\begin{proof}

Considering the Markov chain $v_{0,r} \longrightarrow m_r \stackrel{D}{\longrightarrow} \bar{v}_{0,r}$ and 
using the Data Processing Theorem,
\begin{equation}
    I(m_r,\bar{v}_{0,r})<I(m_r,v_{0,r})
    \label{eq:data_ineq}
\end{equation}
For neural network $D$ that estimates jointly all of the segments we have from Lemma ~\ref{lemma:joint_fisher},
\begin{equation}
    J(\bar{v}_{0}) = \frac{L}{w}\cdot J(\bar{v}_{0,r})
    \label{eq:seg_fisher}
\end{equation}
Recall that $\bar{v}_{0}$ is the estimator of $v_0$ from $m$. Referring $v_0$ as $\theta$ in Lemma ~\ref{lemma:fisher_bound} gives,
\begin{equation}
     I(m_r,v_{0,r}) \geq J(\bar{v}_{0,r})
    \label{eq:fisher_ineq}
\end{equation}
Plugging Eq.\ref{eq:fisher_ineq}, and Eq.\ref{eq:seg_fisher} gives,
\begin{equation}
    \frac{L}{w} \cdot I(m_r,v_{0,r}) \geq J(\bar{v}_{0}) 
\end{equation}
\end{proof}

Based on this lemma, we obtain the following bound.
\begin{theorem}[SDR upper bound]
\label{thm:main}

Let $m_r$, $v_{0,r}$, $L$, $w$ be defined as previously stated.
For any neural network, the SDR upper bound that separates a mixture of $C$ speakers is given by:
\begin{align}
    SDR(v_{0},\bar{v}_{0}) & \leq 10log_{10}\left (\frac{L}{w}\cdot Var(v_0)\cdot I(m_r,v_{0,r})\right )
    \label{eq:final_bound}
\end{align}
\label{thrm:uppr_bnd}
\end{theorem}
\begin{proof}

The Cramer Rao Lower Bound \cite{Rao1992} states that for any unbiased estimator there is a lower bound on the estimator square error i.e. for estimator of $\bar{v}_{0,r}$
\begin{equation}
    Var(\bar{v}_{0,r}) \geq \frac{1}{J(\bar{v}_{0,r})}
    \label{eq:crb_v}
\end{equation}

Applying Eq.~\ref{eq:crb_v} to Definition.~\ref{defintion:sdr} gives an upper bound for estimating the single segment:
\begin{dmath}
    SDR(v_{0,r},\bar{v}_{0,r}) \leq 10log10(Var(v_{0,r})J(\bar{v}_{0,r}))
    \label{sdr_term}
\end{dmath}
Using the upper bound for the Fisher information Lemma.\ref{lemma:fisher_bound} yields:
\begin{equation}
    SDR(v_{0,r},\bar{v}_{0,r}) \leq 10log10(Var(v_{0,r})I(m_r,\bar{v}_{0,r}))
\end{equation}
The data processing inequality states the the mutual information is reduced after each processing, thus we obtain
the following:
\begin{equation}
    SDR(v_{0,r},\bar{v}_{0,r}) \leq 10log10(Var(v_{0,r})I(m_r,v_{0,r}))
    \label{eq:almost_bound}
\end{equation}
Using Lemma.\ref{lemma:joint_fisher} gives the upper bound:
\begin{equation}
    SDR(v_{0},\bar{v}_{0}) \leq 10log10\left (\frac{L}{w}\cdot Var(v_{0})I(m_r,v_{0,r})\right )
\end{equation}

\end{proof}
Theorem~\ref{thrm:uppr_bnd} states that the upper bound for separation algorithm, is a function of: (i) the number of speakers $C$, (ii) the number of segments that jointly estimated $\frac{L}{w}$, (iii) the mutual information $I(m_r,v_{0,r})$. 

Specifically, the bound is linear with $L$, the length of the sequence. Furthermore, it is inversely proportional to the segment length $w$. There is a trade-off between $w$ and the lowest frequencies in each segment, thus, decreasing $w$ also decreases $Var(v_{0,r})$.

A clear limitation of the bound is that it holds only for a network $D$ that jointly process i.i.d stationary segments. This is not the case if a neural network processes the entire signal without segmentation. However, current architectures, including the attention architecture, tend to be myopic.

\section{Method}
\label{sec:method}
Our method, denoted as as SepIt, consists of sequential processing of the estimated signals, where each iteration contains a replica of the basic model.
Consider again the mixture signal, $m$, consisting of $C$ speakers and of length $L$.
First, a backbone network, $B$, is used for initial separation estimation. The backbone can change between different data sets, and it does not learn during the training phase of the SepIt model. 
Expanding the notation in Sec.~\ref{sec:upper_bound}, denote the estimation of the $i$-th speaker in the $j$-th iteration as $\bar{v}_{i}^j$. Collectively, we denote the list of obtained speakers as $\bar{{v}}^j = [\bar{v}_0^j,\bar{v}_1^j,\dots, \bar{v}_{C-1}^j]$. The output of the backbone, $B$, is $\bar{{v}^0}$.
\begin{equation}
    \bar{{v}^0} = B(m)
\end{equation}
Where $\bar{{v}} \in \mathbb{R}^{C\times L}$ is all speaker estimation.

Both the current estimation, $\bar{{v}}^{j-1}$, and the mixture, $m$, pass through an encoder, $E$, of a 1-D convolution with $N$ channels, kernel size $K$ and stride $\frac{K}{2}$, followed by a ReLU activation function:
\begin{dmath}
    \Acute{m} = E(m)
\end{dmath}
\begin{dmath}
    \Acute{v}^{j-1} = E(\bar{v}^{j-1})
\end{dmath}
\begin{figure}[b]
    \centering
     \includegraphics[width=1\linewidth]{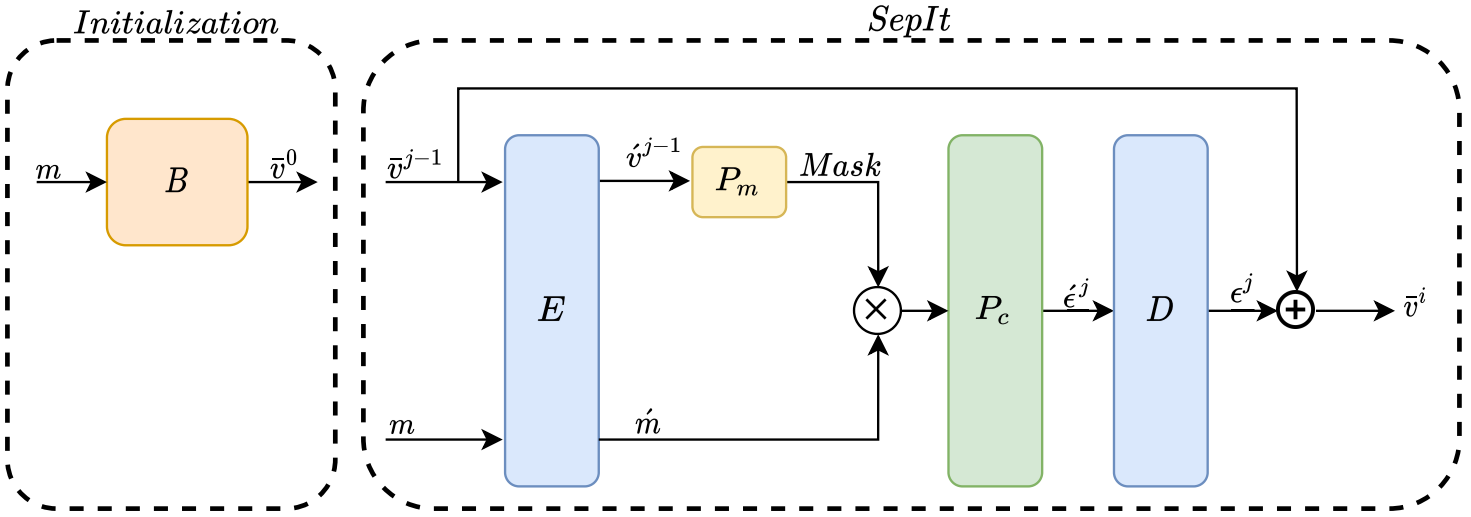}
    \caption{The SepIt block architecture.
    }
    \label{fig:sepitblock}
\end{figure}
Where $\Acute{m}\in \mathbb{R}^{N\times L'} ,\Acute{v}^{j-1}\in \mathbb{R}^{C\times N \times L'} $  are the latent space for the mixture and current speech estimations, respectively, and $L' = \frac{L}{K}$ is the strided latent dimension.
The encoder of $m$ shares its weights with the encoder of $\bar{v}^{j-1}$.

The latent estimation $\Acute{v}^{j-1}$ is treated as a prior on the latent masking. A deep neural network, $P_m$, is used to evaluate a latent space mask.
$P_m$ consists of 3 ResBlocks, each a 1-D version of the Residual block from ResNet, each with $N$ channels, kernel size of 3, and ReLU activation function. 
The latent mask is multiplied element-wise by the mixture latent representation. 
To subtract residuals from other sources, a $1\times 1$ Convolution layer $P_c$ with $N\times C$ channels is then introduced. The procedure is:
\begin{equation}
    \Acute{\epsilon}^{j} =P_c( P_m(\Acute{v}^{j-1})\circ \Acute{m})\,,
\end{equation}
where $\Acute{\epsilon}^{j}\in \mathbb{R}^{C\times N\times L'}$ is the latent space residual, and $\circ$ is the element-wise product operator.
The latent residual $\Acute{\epsilon}$ is passed to a decoder layer, $D$. Later, a skip-connection is introduced, such that the $j$-th iteration estimation is the summation of the decoded $\epsilon$ and the previous estimation,
\begin{equation}
    \bar{y}^j = \bar{y}^{j-1} + D(\Acute{\epsilon}^j)
\end{equation}
A single SepIt block is depicted in Fig~\ref{fig:sepitblock}.
The training algorithm is depicted in Alg.~\ref{alg:diffalg}. The algorithm starts with the initialization of the threshold $u$ to 1. The zero-th iteration is set using the backbone model as in line~\ref{lst1}. Next, the algorithm evaluates the SepIt neural network over the previous iteration, and takes a gradient descent step over the loss function. The threshold $u$ is updated by the difference between $I(m,\bar{v}^j)$ and $I(m,\bar{v}^{j-1})$. 

The algorithm proceeds to apply the SepIt neural network until the maximum number of iterations is reached or the threshold is lower or equal to zero. This stopping criterion follows the derivation of Thm.\ref{thrm:uppr_bnd}, which indicates that the mutual information between the speaker estimation $\bar{v}^j$ and the mixture $m$ can indicates the maximal achievable SDR. Thus, a per-sample stopping criterion is derived when $I(m,\bar{v}^j)$ stops increasing relative to previous iteration $I(m,\bar{v}^{j-1})$. This stopping criterion can be computed also on the test-set samples since it does not require the ground truth.

\textbf{Loss Function} Following \cite{roux2018sdr}, a Scale Invariant Signal to Distortion Ratio (SI-SDR) is used as the loss function, which is a similar loss to SDR, but demonstrated better convergence rate. The overall loss function is given by:
\begin{equation}
    {\mathcal{L}}(v,\bar{v}) = 10log10\left(\frac{||\Tilde{v}||^2}{||\Tilde{e}||^2}\right )
\end{equation}
where $\bar{v}$ is the estimated source, $\Tilde{v}=\frac{<v,\bar{v}>v}{||v||^2}$, and $\Tilde{e}=\bar{v}-\Tilde{v}$.

\begin{algorithm}[h]
\caption{The SepIt algorithm. \\ \textbf{Input:} $m$ - mixture signal, $C$ - number of speakers, $B$ - Backbone network, MAXITER - maximum number of iterations \\
\textbf{Output:} $\bar{v}^{j}$ - estimated separated speakers.}
\label{alg:diffalg}
\begin{algorithmic}[1]
\STATE $\bar{v}^{0}\gets B(m)$  
\STATE $j \gets 0$ 
\STATE $u \gets 1$ \label{lst1}
\STATE $l \gets 0$
\WHILE{$u \geq 0$ and $j \leq MAXITER$} \label{lst:line:blah2}
\STATE $\bar{v}^{j} \gets SepIt_{j}(\bar{v}^{j-1},m)$ 
\STATE $l \gets SDR(\bar{v}^{j},v)$ 
\STATE Take gradient descent step on $-l$
\STATE $j \gets j+1$
\STATE $u \gets I(m,\bar{v}^j)-I(m,\bar{v}^{j-1})$
\ENDWHILE\label{diffsepdwhile}
\STATE \textbf{return} $\bar{v}^{j}$
\end{algorithmic}
\end{algorithm}

\section{Experiments}
\label{sec:exp}

\begin{figure}[h]
     \includegraphics[width=0.85\linewidth,height=0.6\linewidth]{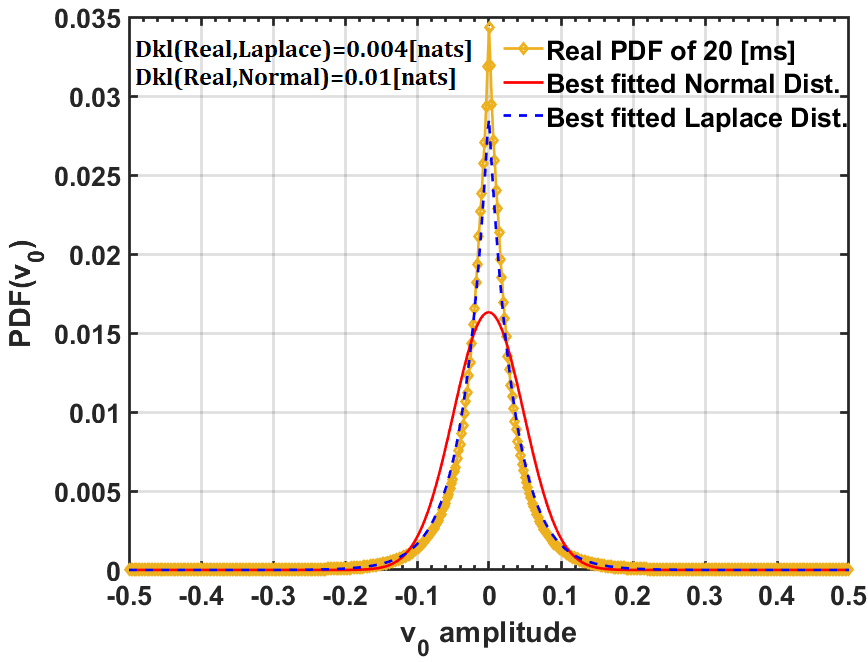}
    \caption{Distribution comparison of empirical PDF, best fitted normal and Laplace distribution. X axis - the signal sample amplitude, Y-axis the PDF of the sample amplitude. The Laplace distribution is the best fit to the empirical PDF}
    \label{fig:dist_compare}
\end{figure}

\noindent{\bf Validating Assumption \ref{asmp:laplace}\quad} This assumption is extensively validated by \cite{Jensen}, yet we re-evaluated it again over the LibriMix training set. The LibriMix dataset containing over $100$ hours of speech is investigated. Each speaker signal is split into different non-overlapping segments, each $20[ms]$ long. Fig.~\ref{fig:dist_compare} compares the empirical PDF with the best fitted Laplace distribution and normal distribution. Evidently, the Laplacian distribution provides a much better fit, The Kullback–Leibler that presented in the figure supports this conclusion too. 

\begin{figure*}[t]
    \centering
     \includegraphics[width=0.8\linewidth,height=0.38\linewidth]{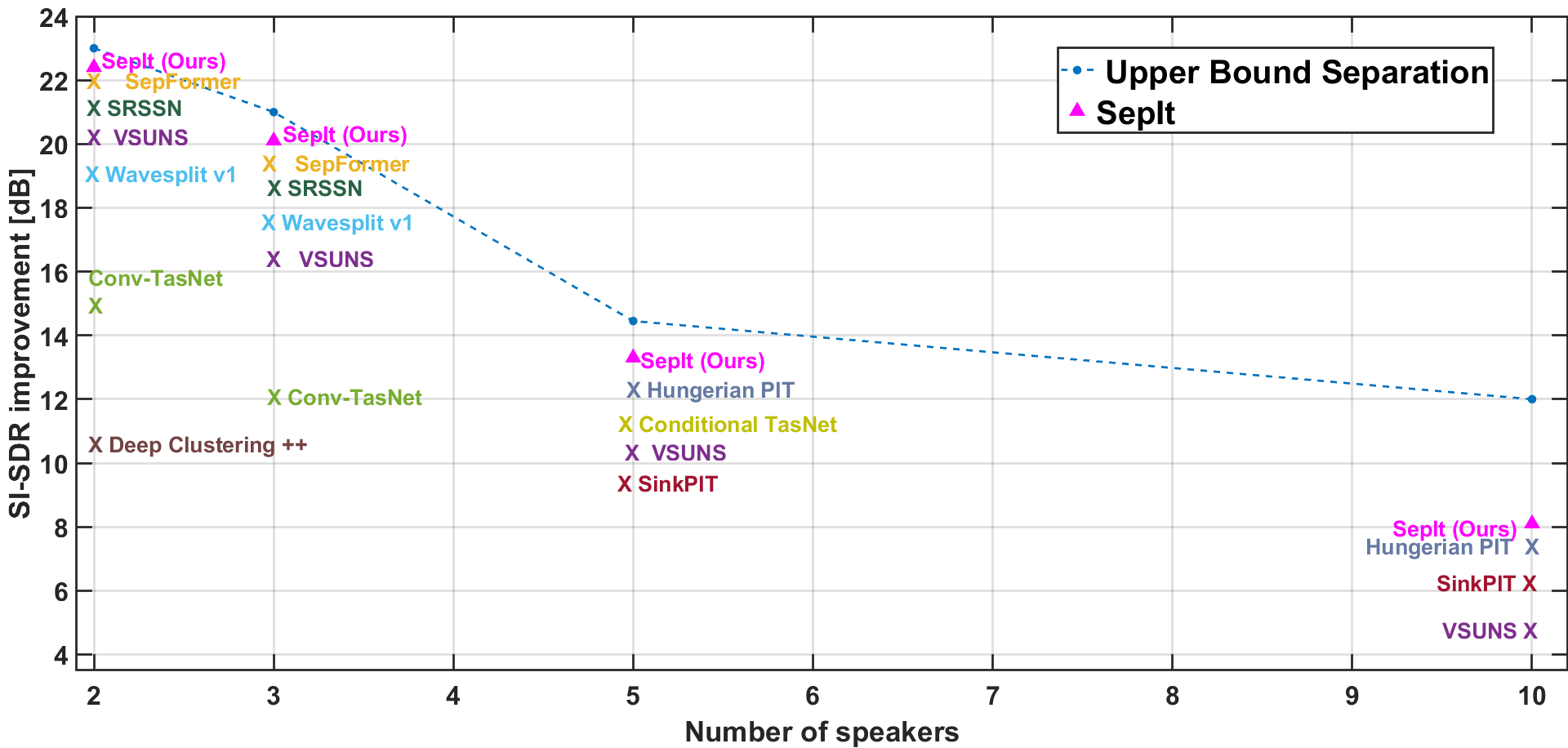}
    \caption{The upper bound for single channel source separation (blue curve).
   SepIt, denoted by a pink triangle, produces state-of-the-art results.}
    \label{fig:sdr_res}
\end{figure*}

\noindent\textbf{Separation results\quad}
For all experiments, the Adam \cite{kingma2017adam} optimizer is used with a learning rate of $5\cdot10^{-4}$ and a decay factor of 0.95 every epoch. The window size $w = 20[ms]$ and speech length $L=4[s]$. Other hyperparameters are summarized in Tab.~\ref{tab:model_size}.
The experiments are divided into two categories based on their Backbone architecture: (i) Transformer -based and (ii) LSTM-based. Each backbone is used in the range for which it is currently the state of the art.
We use dynamic mixing augmentation, introduced in \cite{zeghidour2020wavesplit}, which consist of creating new mixtures in run time, noted as \textbf{DM}.

\begin{figure}
 \centering
\begin{tabular}{c}
 \includegraphics[width=0.85\linewidth]{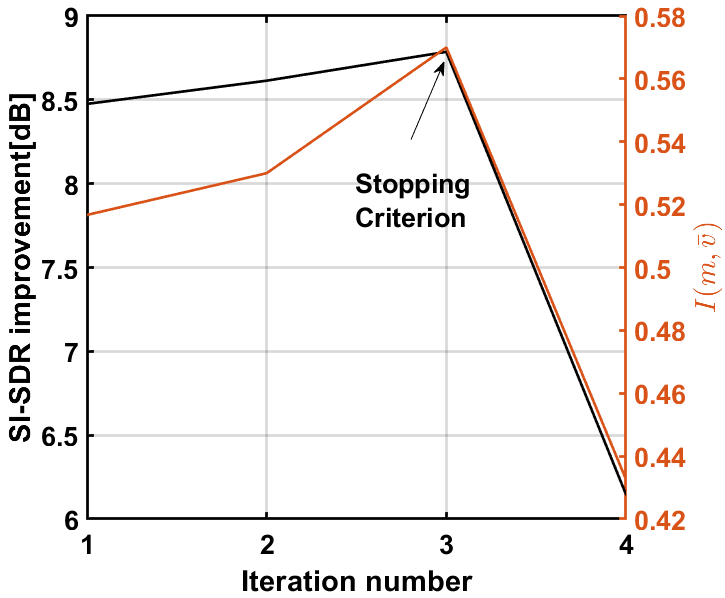}
 \end{tabular}
 \vspace{-1mm}
 \caption{SDR improvement for $5$ speakers and the Mutual information between the mixture and the estimated speaker.}
\label{fig:iter_stop}
\end{figure}

\begin{table}[h!]
\caption{Hyperparameters for the different number of speakers.}
    \label{tab:model_size}
\centering
 \begin{tabular}{@{}l@{~}c@{~}c@{~}c@{~}|l@{~}c@{~}c@{~}c@{}} 
 \toprule
No. speakers &  N & K & Params[M]& No. speakers &  N & K & Params\\
\midrule
2 & 256& 4 & 4.6& 5 & 128 & 4& 1.95\\
\midrule
3 & 256& 4 & 7.2&10 & 128& 4 & 2.7\\
 \bottomrule
 \vspace{-7mm}
\end{tabular}

\end{table}
For all experiments, 5 iterations are conducted, where \textbf{SC} implies that the stopping criterion per sample is activated. Fig.~\ref{fig:iter_stop} depicts the behavior of the criterion on  a typical sample from the test set.

\textbf{Transformer Backbone}
As seen from the upper bound, the current state-of-the-art results are close to the bound. Thus, only a small improvement is expected when experimenting above it with SepIt. The results using a current state-of-the-art Transformer network \cite{subakan2021attention} as $B$ over the Wall Street Journal Mix $2/3$ (WSJ0-2/3 Mix) are summarized in Tab.~\ref{tab:results_2_3spkrs}.

\begin{table}[h!]
\caption{The performance obtained by our method, as well as the Transformer backbone network. See text for details.}
    \label{tab:results_2_3spkrs}
\centering
 \begin{tabular}{@{}l@{~}cc@{}} 
 \toprule
& \multicolumn{2}{c}{SI-SDRi $[dB]${$(\uparrow)$}}\\
 \cmidrule(lr){2-3}
Method &  WSJ0-2Mix  & WSJ0-3Mix \\
\midrule
Upper Bound & 23.1 & 21.2\\
 \midrule
(i) SepFormer &20.4 & 17.6\\
(ii) SepFormer + DM &22.3 & 19.8\\
\midrule
{\bf(iii) SepIt + SepFormer + DM} &{\bf22.4} & {\bf20.1}\\
{\bf(iv) SepIt + SepFormer + DM + SC} &{\bf22.4} & {\bf20.1}\\
 \bottomrule
 \vspace{-7mm}
\end{tabular}

\end{table}

First, we observe that the current state of the art is approaching the upper bound. For 2 and 3 speakers, the gap between the upper bound and current results is 0.8dB and 1.4dB, respectively. Second, as expected, our SepIt model is able to improve only slightly the result with $2$ speakers, with an improvement of $0.1dB$ over the current state of the art, while for $3$ speakers, SepIt improves the result by 0.3dB. We note that for 2/3 speakers only few examples had early stopping with the stopping criterion ({\bf SC}), which resulted in similar results.

\textbf{LSTM Backbone}
In LSTM-based architectures the state-of-the-art network for a large number of speakers was provided by \cite{dovrat2021manyspeakers}. We apply the SepIt network over the Libri5Mix and Libri10Mix datasets. The results are summarized in Tab.~\ref{tab:results_5_10spkrs}. As expected, here the improvement over the current state-of-the-art architecture is more prominent than in the Transformer case, where for $5$ speakers, SepIt improves results by 1.0dB, and for $10$ speakers by 0.5dB.
\begin{table}[h!]
\caption{The performance obtained by our method, as well as the LSTM backbone network. See text for details.}
\label{tab:results_5_10spkrs}

\centering
 \begin{tabular}{@{}l@{~}cc@{}} 
 \toprule
& \multicolumn{2}{c}{SI-SDRi $[dB]$ {$(\uparrow)$}}\\
 \cmidrule(lr){2-3}
Method &  Libri5Mix  & Libri10Mix \\
\midrule
Upper Bound & 14.5 & 12.0\\
Gated LSTM~\cite{dovrat2021manyspeakers} &12.7 & 7.7\\
SepIt+ Gated LSTM &13.2 & 8.0\\
SepIt+ Gated LSTM + DM & 13.6 & 8.1\\
\textbf{SepIt+ Gated LSTM + DM +SC} &{\bf13.7} & {\bf8.2}\\
 \bottomrule
 \vspace{-12.5mm}
\end{tabular}

\end{table}

\section{Summary}
In this work, a general upper bound for the Single Channel Speech Separation (SCSS) problem is derived. The upper bound is obtained using the Cramer-Rao bound, with fundamental modeling of the speech signal.
We show that the gap between the current state-of-the-art network for 2 and 3 speakers and the obtained bound is relatively small, 0.8[dB] and 1.4[dB] respectively. This may indicate that future research should focus more on reducing the model size or the required training set and less on improving overall accuracy. The gap for 5 and 10 speakers is larger, and there is still room for substantial improvement in separation accuracy. Using the upper bound, a new neural network named SepIt is introduced. SepIt takes a backbone network estimation, and iteratively improves the estimation, until a stopping criterion based on the upper bound is met. SepIt is shown to outperform current state-of-the-art networks for 2, 3, 5 and 10 speakers. 


\section{Acknowledgments}
The contribution of Shahar Lutati is part of a Ph.D. thesis
research conducted at Tel Aviv University.
This project has received funding from the European Research Council (ERC) under the European Unions Horizon 2020 research and innovation programme (grant ERC CoG 725974). 

\bibliography{mybib}
\bibliographystyle{IEEEtran}


\end{document}